\documentstyle[11pt]{report}

\def\ZzZ{{\hbox{\tenrm Z\kern-.31em{Z}}}}
\def\CcC{{\hbox{\tenrm C\kern-.45em{\vrule height.67em width0.08em depth-
.04em
\hskip.45em }}}}

\newcommand{\lab}{\label}

\newcommand{\bc}{\begin{center}}
\newcommand{\ec}{\end{center}}
\newcommand{\be}{\begin{equation}}
\newcommand{\ee}{\end{equation}}
\newcommand{\bea}{\begin{eqnarray}}
\newcommand{\eea}{\end{eqnarray}}
\newcommand{\bs}{\begin{subequations}}
\newcommand{\es}{\end{subequations}}
\newcommand{\beq}{\begin{eqalignno}}
\newcommand{\eeq}{\end{eqalignno}}

\def\lab{\label}

\begin{document}

$$  $$

\bc 
{\Large \bf The decoherence criterion}

\vspace{.5cm}

{Eleonora Alfinito${}^{a}$\footnote{ alfinito@sa.infn.it}}, 
Rosario G. Viglione${}^{b}$\footnote{ ros@alex.chem.unisa.it  } and
Giuseppe Vitiello${}^{a}$\footnote{vitiello@sa.infn.it}

\vspace{0.5cm}

${}^{a}${\it Dipartimento di Fisica, Universit\`a di Salerno, 84100}


{\it and INFN Gruppo Collegato di Salerno and INFM Unit\`a di Salerno}


${}^{b}$ {\it Dipartimento di Chimica, Universit\`a di Salerno, 84100}

\ec

\vspace{1cm}

\smallskip
$$ $$

{\bf Abstract}

\hspace{0.5cm} The decoherence mechanism signals the limits
beyond which the system dynamics approaches the classical behavior.
We show that in 
some cases decoherence may also signal the limits beyond which 
the system dynamics has to be described by quantum field theory, rather
than by quantum mechanics.

\vspace{0.5cm}
\smallskip


$$  $$


The problem of the range of applicability of quantum mechanics (QM) has 
been
the object of much attention and studies since the early days of its
foundation. However, only at a later time it has been fully recognized and
appreciated the purely quantum phenomenon of decoherence \cite
{zhe,zurek,zurek2,joos,giulini}, which in many cases signals the {\it 
appearance
of a classical world in quantum theory} \cite{giulini}. Decoherence is 
normally
triggered by the interaction of the system with the environment and 
formally
consists in suppressing the off-diagonal elements of the reduced density
matrix. Its effect is such that quantum superposition in the system
wave function is destroyed and thus, provided the time-scale ${\tau}_{dyn}$
characteristic of the dynamics is much greater than the decoherence
time-scale ${\tau}_{dec}$, ${\tau}_{dyn} \gg {\tau}_{dec}$, the classical
behavior may be approached. It should be noted that sometimes the system 
under
consideration, assumed to be a closed system, is viewed as composed of two
subsystems and we are actually interested only in one of them. The other
subsystem is then regarded to be the environment in which the former one is
embedded. The reduced density matrix is the one obtained by tracing over 
the degrees of freedom of the environment (the rest of the Universe, 
whatever it is).

Decoherence provides an interesting explanation why we do not experience
superpositions of objects in the macroscopic world. In fact interaction 
with
the environment produces decoherence in the superposition of 
macroscopically
separate positions so that the familiar classical behavior of
non-superposing macroscopic objects is obtained. However, we also have
experience (and sound theoretical understanding) of the existence of {\it 
macroscopic quantum systems}, such as superconductors, ferromagnets,
crystals, and in general systems presenting ordered patterns, where
coherence over macroscopically large distances appears to be particularly
stable against environment perturbations. Any system is made by quantum 
components. It is, of course,
not in such a trivial sense that macroscopic systems such as a
superconductor or a crystal are quantum systems. These are quantum systems
in the specific, non-trivial sense that their macroscopic (classical)
behavior cannot be explained without recourse to quantum theory. It is then
natural, and necessary, asking the question of the consistency between the
decoherence phenomenon and the existence of macroscopic quantum systems.
This is indeed the question we address in this paper.

Our conclusion will be that the decoherence mechanism characterizes QM by
designing its borderlines also with quantum field theory (QFT), besides, 
as already known, with classical mechanics. In other words, our result 
suggests the use of the decoherence mechanism as an useful {\it
criterion} 
to scan the border QM/QFT. 

For the sake of clarity it might be convenient to start our discussion by 
shortly reviewing some general aspects of the QFT approach to the study 
of macroscopic quantum systems, such as 
superconductors, ferromagnets, crystals, etc. These systems have the  
common feature of presenting some kind of ordered patterns in the ground 
state. The observable describing such an ordering is called the order 
parameter and it is a macroscopic observable in the sense that it is not 
strongly affected by quantum fluctuations. The spontaneous breakdown of 
symmetry in QFT provides the mechanism out of which the ordered patterns 
are dynamically generated.

Suppose that the system Lagrangian (or the field equations from it
derived) is invariant under some group G of symmetry
transformations. The symmetry is said to be spontaneously broken when
the lowest energy state (the vacuum or ground state) of the 
physical system is not invariant under
the same symmetry group G, but under one of its subgroups. For example,
for a metal, which below some critical temperature may exhibit
superconductivity, the Lagrangian is invariant under the U(1)
transformations. However, the ground state of the metal in the
superconducting phase is not invariant under the U(1)
transformations. The ground state of the same metal above the critical
temperature, in the "normal" (i.e. non superconducting) phase, is, on
the contrary, U(1) symmetric. The Lagrangian for a gas of particles
endowed with magnetic moment, which under specific boundary conditions 
may be found in the ferromagnetic
phase, is invariant under the SU(2) group of rotations. However, the
system in the ferromagnetic phase, below the Curie temperature, is
invariant under the U(1) subgroup of the original SU(2).  The Lagrangian
for a gas of atoms or molecules, which below a critical temperature may
be arranged in a crystal phase, is invariant under the continuous space
translations. However, the same system of atoms or molecules when it is 
in the crystal phase has the lowest energy state which is not 
invariant under continuous space translations, but under 
discrete translations of the length of the lattice spacing or
of its multiples.  Whenever, the symmetry of the Lagrangian is not the
same as the symmetry of the vacuum we have spontaneous symmetry
breakdown. The dynamics of a system, given in terms of the
Lagrangian, may thus generate at the observable level, under different
boundary conditions (e.g. different temperatures), different phases of
the system (the normal phase and the superconducting phase, the
ferromagnetic phase and the  non-ferromagnetic one, the crystal phase
and the amorphous one). The crystal ordering, the superconductor
ordering, etc., thus appears in the non-symmetric (or "less" symmetric)
ground state: order is in fact lack of symmetry.

The Goldstone theorem \cite{itzikson} states that whenever a continuous
symmetry is spontaneously broken the dynamics implies the existence of
gapless (massless) modes: they are called the Nambu-Goldstone (NG) boson
particles or quanta. We will not analyze the case in which there is a
gauge field in the theory, which would not change the substance of our
conclusions. Also, we will not report the derivation of the Goldstone
theorem since it can be easily found in any standard textbook of QFT
(see e.g. \cite{itzikson}).

The Nambu-Goldstone bosons are the quanta of long range correlation
among the system constituents. Examples of NG boson quanta are the
phonons in the crystals, the magnons in the ferromagnets, the Cooper
pairs in the superconductors. As already said, these quanta are of
dynamical origin, namely they are not found in the symmetric or normal
phases (which are typically obtained above some critical
temperature). The NG bosons are thus collective modes, the long range 
correlation, of which they are the quanta, is responsible for the
ordering in the ground state of the non-symmetric (ordered) phase: the
NG quanta are the carriers of the ordering information throughout the
system. It is in this way that the order is generated by the quantum
dynamics and appears as a macroscopic, diffused property of the system
in the non-symmetric phase. In other words, in the ordered phase the
system components get, so to say, "trapped" by the long range
correlation, they cannot behave as individual particles. Some of their
degrees of freedom get frozen by the NG long range correlation, which is
a dynamical consequence of the "lack" of symmetry, and this manifests
itself as the system macroscopically observable ordered patterns. We
thus arrive at the understanding of the existence of macroscopic quantum
systems. All of this is formally expressed in terms of coherent
condensation of NG boson quanta in the ground state
\cite{itzikson,Anderson,Umezawa,Um}. Since these quanta are massless
their coherent condensation in the lowest energy mode does not add
energy to the ground state and the observed high stability of the
ordered phases is thus explained. The mathematical formalism of the
many-body theory of superconductors, ferromagnets, crystals and other
macroscopic quantum systems is given in great details in standard
textbooks \cite{itzikson,Anderson,Umezawa} and therefore we do not
report it here.

A given system thus may possesses degenerate ground states, each
corresponding to a physically different phase in which the system may be
observed under different boundary conditions. These degenerate ground
states are in fact unitarily inequivalent, i.e. it does not exist any
unitary operator transforming one of them into another one. In
different words, one cannot express the ground state of a specific phase
in terms of the ground state of another, different phase: the crystal
ground state cannot be expressed in terms of the amorphous ground state,
the superconducting ground state cannot be expressed in terms of the
normal one, and so on. As well known \cite{itzikson} - 
\cite{Bratteli}, this is formally
expressed, e.g. for the superconducting Bogoliubov or $BCS$ ground state
$|0(BCS)>$, by
\be\lab{bog} 
<0|0(BCS)> \rightarrow 0 ~ ~~~for~~~ V \rightarrow \infty ~,
\ee 
in the infinite number of degrees of freedom limit (infinite volume
$V$ limit).   Here $|0>$ denotes the "normal" ground state (i.e. the
non-superconducting phase ground state), and $|0(BCS)>$ is formally
expressed at finite volume as $|0(BCS)> \equiv 
\Pi_{k} (U_{k} + V_{k}c_{{\bf k}+}^{\dag}c_{{- \bf k}-}^{\dag})|0>$, 
with 
$U_{k}^{2} + V_{k}^{2} = 1$ and $c_{{\bf k}\pm}^{\dag}$ denoting the
creation operators for the spin up and down electron, respectively
\cite{Rickayzen}. The meaning of Eq. (\ref{bog})
is that in the infinite volume limit 
the ordered ground state $|0(BCS)>$ cannot be reached by
building up perturbations around the non-ordered one $|0>$: the state
$|0(BCS)>$ cannot be expressed in terms of $|0>$, and vice-versa. We
have a typical non-perturbative phenomenon: the different physical
phases are described by states belonging respectively to different 
Hilbert spaces, each of them corresponding to a
specific ground state. These physically different spaces are the
unitarily inequivalent representations of the canonical commutation
relations and, for the above example, we denote them by
$\{|0>\}$ and $\{|0(BCS)>\}$.  Eq. (\ref{bog}) is
a typical QFT result: in fact (\ref{bog}), which states the unitary
inequivalence between the representations $\{|0>\}$ and $\{|0(BCS)>\}$,
cannot hold in QM where the von Neumann theorem states
that {\it all} the representations are unitarily, and thus physically,
equivalent \cite{Bogoliubov,Sewell,Bratteli,QM}.   On
the contrary, physically different phases are allowed in QFT, 
since the von
Neumann theorem does not hold in QFT and there exist infinitely many
unitarily inequivalent representations of the canonical commutation
relations. 

Eq. (\ref{bog}) thus states that the systems
presenting different physical phases cannot be described purely 
in terms of QM. QM is not enough "rich" to allow
unitarily inequivalent representations, as rigorously implied by the von
Neuman theorem. In this paper we are going indeed to confirm such a
result from the perspective of the decoherence mechanism in QM. 

We will
consider as an example the commonly observed process of the formation of
a crystal out of an ion solution. The result implied by QM decoherence
formulas would prevent the formation of the crystal, since the computed
decoherence time for the ions in the solution is
too much short with respect to the time one has to wait before the
crystal gets formed in common observations.   This does not mean that
those QM formulas are wrong, neither, of course, it means that our
system is a classical one. It means that it is wrong to apply these QM
formulas to the ions in solution in order to study the process of the
crystal formation. Thus, in QM, once one starts with the ion solution
phase, dechoerence tells us that no other phase (the crystal phase) 
is reachable, which indeed is in perfect agreement with 
what the von Neumann theorem states.

Let us now turn our attention to the QM decoherence
scenario.   There is a large body of literature on decoherence and for
the sake of  shortness we do not reproduce here the already published
derivations of the  formulas we are going to use. We refer for that to
the quoted papers.    To be specific, we focus on the formation of,
e.g., the binary crystals listed in Table 1. We stress that the system
we study is {\it not} the {\it already formed} crystal, but the solution
of ions out of which one expects (observes) the crystal will be
formed. Thus, according to the usual chemical recipe, we consider a
solution (typically, the water is the solvent) of the constituent
elements (e.g. a solution of $Na^{+}$ ions and $Cl^{-}$ ions) and wait,
in specific conditions of temperature, density, etc., till the
crystallization occurs. This happens when the saturation of the
solution is reached. At the crystallization point, the saturation
concentrations can be quite different in different cases, depending on
the crystal one wants to obtain, ranging, for example, from $1$ ion of
$K^{+}$ for $4$ molecules of water for $KF$, or $1$ ion of $Na^{+}$ for
$10$ molecules of water for $NaCl$, to $1$ ion of $Ag^{+}$ for $10^{8}$
molecules of water for $AgBr$, till $1$ ion of $Pb^{+}$ for $10^{15}$
molecules of water for $PbS$ \cite{concentrazioni}.  The ions in the
solution are normally bound, due to Coulomb forces, to water molecules,
for example the $Na^{+}$ ion is surrounded by four water molecules. The
shielding of the ionic charge by the surrounding water molecules lowers
the intensity of  the Coulomb interaction among ions. Sometimes one adds
a "germ", namely a small crystal of the same kind of the one to be
formed. Such a germ will act as a catalytic structure making more
favorable the aggregation, in the wanted crystal structure, of the ions
in the solution. Sometimes the nucleation is simply produced by some
"defect" or "impurity", e.g., on the walls of the bowl or container of
the solution. One observes the crystal formation in the vicinity of
these defects. At the crystallization point, lowering the temperature of
the solution normally helps the crystal formation, which can occur
within a short lapse of time (from fractions of a second to several
seconds) or in a longer one (from minutes to hours).

The interaction among the ions in the solution is of Coulomb type and we
will consider the ion-ion collisions and the interactions with distant
ions to be two possible sources of decoherence. There are also other
sources of decoherence such as, e.g., the interaction with the
environment (the water in our case), with the crystal germ or with the
defects or impurities, or else with dipole and higher moments of charge
distribution. However, the decoherence effect we compute from ion-ion
collisions and distant ion interactions are so strong that we can
neglect any other decoherence source. Moreover, we will see that the
decoherence time does not strongly depend on the different
concentrations of the different ionic solutions mentioned above.

The density matrix for the single ion may be shown to be proportional to
a function $f({\bf x}, {\bf x}^{\prime}, t)$ which does not depend on 
the ion state \cite{tegmark, zhe}. Here ${\bf x}$ denotes the ion position.
For scattering of environment particles with de Broglie wavelength 
$\lambda$, we have
\begin{equation}\lab{1}
f({\bf x}, {\bf x}^{\prime}, t) = e^{-\Lambda t\left(1-e^{-%
\frac{|{\bf x}^{\prime}- {\bf x}|^2}{2\lambda^2}}\right)} ~,
\end{equation}
that is,
\begin{equation}
f({\bf x}, {\bf x}^{\prime}, t)\sim\left\{ 
\begin{array}{ll}
e^{-\Lambda t \frac{|x^{\prime}-x|^2}{2\lambda^2}} & 
\mbox{ for  $|{\bf x}'-
{\bf x}|\ll\lambda$} \\ 
e^{-\Lambda t} & \mbox{ for
$|{\bf x}'-{\bf x}|\gg\lambda$}
\end{array}
\right.,
\end{equation}
where $\Lambda \equiv n <\sigma v>$ is the scattering rate cross section.
The product of the cross section $\sigma$ by the velocity $v$ is averaged
over the thermal velocity distribution at $T = 310~ K$, $n$ is the density
of the scatterer centers (ions). 
Eq. (3) implies that for the 
single ion  the spatial superposition decays exponentially 
as $\Lambda ^{-1}$ and that for $N$ ions as $(\Lambda N)^{-1}$.
 Let us consider the case of $Na^{+}$ 
and $Cl^{-}$ ions solution. The de Broglie wavelength of the 
$Na^{+}$ ion (mass $m_{Na} = 22.990~ amu$) or the $Cl^{-}$ ion (mass 
$m_{Cl}
= 35.453 ~ amu$) is
\begin{equation}
\lambda=\frac{2\pi\hbar}{\sqrt{3mkT}} \sim 0.3 \AA ~.
\end{equation}
In the present case of high ionic concentration, 
the smallest acceptable inter-ionic distance ($|{\bf x}-{\bf
x}^{\prime}|$) 
cannot be of course smaller than the one corresponding to the 
elementary cell of the crystalline salt to be formed, which typically is
of 
the order of few
\AA. Even for such a distance we are clearly in 
the $|{\bf x}-{\bf x}^{\prime}| \gg \lambda$ limit of
equation (2). (Such a condition is certainly satisfied also in the case
of 
the other, smaller concentrations of the other considered ionic
solutions). 
In the case of two 
ion-ion collision with unit charge $q_{e}$
and with relative velocity $v$ the cross section is evaluated to be \cite
{tegmark}
\begin{equation}
\sigma \sim \left( \frac{g{q_e}^{2}}{m {v}^{2}} \right)^{2} ~,
\end{equation}
where $g =1/4\pi \epsilon_{0}$ is the Coulomb constant. The decoherence 
time 
${\tau}_{1} \equiv ({\Lambda N})^{-1}$ for $N$ ions
is then given by
\begin{equation}
\tau_{1} \sim \frac{\sqrt{m(kT)^{3}}}{Nng^2q_{e}^4} ~.
\end{equation}
We have used $v \sim \sqrt{\frac {kT}{m}}$ at thermal equilibrium. 
$N$ is taken to be of the order of $10^{23}$ (about a mole of ions), 
and for $Na^{+}$ and $Cl^{-}$ at the crystallization point, it is 
$n=2.163$ ~x~ $10^3 kg~ m^{-3}$
\cite{dati}. 
Then we have:
\begin{equation}
\tau_{1} \sim\frac{\sqrt{m(kT)^{3}}}{10^{23}\frac{2163}{m_{Na} + 
m_{Cl}}g^2q_{e}^4} \sim 4.6 ~10^{-40} ~s.
\end{equation}

The order of magnitude of this result does not change much even for the 
other ionic solutions, which confirms the correcteness of the used
approximations. The $\tau_{1}$ values we obtain for the other ionic
solutions are reported in Table 1.  

In a similar way, one can show that the decoherence 
time $\tau_{2}$ for the interaction with distant ions 
is given by \cite{tegmark}:
\begin{equation}\lab{7}
\tau_{2} \sim\frac{\sqrt{m(kT)}}{Nnag q_{e}^2}~.
\end{equation}
In this case the smallest acceptable distance between
two $Na^{+}$ ions is $a = 5.64~ \AA$ \cite{dati} (the edge of the cubic 
elementary cell).

In the following table we report the decoherence time for ion-ion collisions 
($\tau_{1}$) and for interactions with distant ions ($\tau_{2}$) in the
ionic solutions for the formation of a set of 
crystalline binary compounds. 

$$  $$

\noindent Table 1. Decoherence time for ion-ion collisions $(\tau_{1})$ and
for interactions with distant ions $(\tau_{2})$

\begin{center}
\begin{tabular}{crrrcl}
\hline\hline
\multicolumn{1}{c}{salts} & \multicolumn{1}{c}{$\tau_{1}$/10$^{-40}s$} & 
\multicolumn{1}{c}{$\tau_{2}$/10$^{-38}s$} & \multicolumn{1}{c}{} &  &  \\ 
\hline
NaF & 2.6 & 4.9 &  &  &  \\ 
NaCl & 4.6 & 4.4 &  &  &  \\ 
NaBr & 5.5 & 4.9 &  &  &  \\ 
NaI & 7.1 & 5.8 &  &  &  \\ 
KF & 5.1 & 5.2 &  &  &  \\ 
KCl & 8.3 & 7.1 &  &  &  \\ 
KBr & 9.6 & 7.9 &  &  &  \\ 
CsF & 13.4 & 12.0 &  &  &  \\ 
CsCl & 17.3 & 21.0 &  &  &  \\ 
CsBr & 19.6 & 25.6 &  &  &  \\ 
CsI & 23.6 & 27.8 &  &  &  \\ 
AgCl & 9.5 & 9.2 &  &  &  \\ 
AgBr & 10.7 & 10.0 &  &  &  \\ 
AgI & 15.3 & 12.7 &  &  &  \\ 
ZnS & 7.2 & 7.2 &  &  &  \\ 
PbS & 16.2 & 14.7 &  &  &  \\ \hline\hline
\end{tabular}
\end{center}

$$  $$

According to Table 1, due to the very short decoherence time, there 
would be no 
possibility for the formation of the considered crystals to occur, which
of course contradicts the common experience (in practice the crystal
formation lasts for a time many orders of magnitude longer than
$\tau_{1}$ and $\tau_{2}$ in Table 1).  
We note that in the computation of the decoherence time we have used for
the ions in the solution the shortest possible inter-ionic distances
(and therefore the most favorable to avoid decoherence). We also remark,
that in the case in which the crystal is described as a quantum
mechanical  
n-body system, the wave functions of the constituent ions are centered
at the proper lattice 
sites and present spatial superposition over distances
of the order of the lattice length. Then one could be tempted to
conclude that the above results could apply to the ions which
form the crystals. This conclusion is, however, clearly wrong. The
crystal would not even be stable in this case, in evident contradiction
with common experience.  

The way out of these contradictions is in the fact that the QM
description is not completely adequate 
for the description of the process of the crystal formation 
and for the crystal system: as said in the introductory remarks,  
the crystal
phase transition process and the crystal system are many-body QFT
problems and the binding of the atoms in the crystalline lattice is due
to the long range correlation mediated by the NG bosons (the QFT
formalism can be found in full length and details in standard textbooks,
see e.g. \cite{Umezawa}). In the present case, the decoherence mechanism
thus points to the borderline between QM and QFT. 

The inter-ionic distances used in our computations are of course 
comparable with the wavelength used in the $X$-ray diffraction
experiments with the systems we consider, which is  
$\lambda \sim{1.5\AA}$. This corresponds to oscillation time $\tau_{X}$ 
of the order of $0.5$ x $10^{-18} ~s$, much longer than $\tau_{1,2}$ in
Table 1. 
Suppose, however, one wants to insist in 
using, e.g., Eq. (5). Then, we have
\begin{equation}
\frac{\tau_{1}}{\tau_{X}}=\frac{n_{X}}{n} ~,
\end{equation}
i.e. the $X$-ray typical diffraction time $\tau_{X}$ would 
correspond to an extremely diluted solution where the density $n_{X}$ of
the scatterer centers, say  
$Na$ and $Cl$ ions, is
of the unacceptable order of $10^{-18}$, and, from
\begin{equation}
n_{X}=10^{-18}=\frac{(m_{Na}+m_{Cl})amu}{a^3} ~,
\end{equation}
we have the equally absurd ion spacing of the order of $a\sim10^{-3}m=
10^{7} \AA$.   
Equally unacceptable conclusions are reached by using Eq. (7). 

Again we see that in the case here discussed the
QM decoherence picture and formulas cannot be used. They are not
consistent with familiar experimental 
methodologies. Since these methodologies, on the other hand, confirm the
quantum dynamical nature of the system, the inconsistency points to the
limit of applicability of QM in favor of QFT.   
As far as we know, the existing literature has not paid
attention to this side of the confining limits of QM by using the
decoherence phenomenon.

In conclusion, according to our discussion, decoherence may be then
promoted to the 
relevance of a {\it criterion}, able to discriminate between QM and 
QFT, from one side, and classical mechanics from the
other side. Thus, generally speaking, and contrarily to current common
belief, decoherence does not necessarily 
signals the approaching to the classical mechanics regime; it may also
signals the approaching to the QFT regime, indeed. One must carefully
consider the physics of the system under study in order to correctly
conclude on the implications of decoherence.

This work has been partially supported by INFN, by MURST, by INFM and by
an ESF Network on Topological Defects.

\end{document}